\documentclass[11pt,a4paper]{article}

\usepackage{graphicx}
\usepackage[latin1]{inputenc}

\title{
\textbf{
Dielectric study of the glass transition: correlation with calorimetric data.
}}

\author{
J. A. Diego,
J. Sellar\`es\thanks{e-mail: {\tt jordi.sellares@upc.edu}},
A. Aragoneses,
J. C. Ca{\~n}adas,\\
M. Mudarra and
J. Belana\\
\em\small{Dept. de Física i Enginyeria Nuclear,}\\
\em\small{Universitat Polit\`ecnica de Catalunya (Campus de Terrassa),}\\
\em\small{c. Colom 11, E-08222 Terrassa, Spain.}
}

\date{}

\begin{document}

\maketitle

\begin{abstract}

The glass transition in amorphous poly(ethylene terephthalate) is studied by thermally stimulated depolarization currents (TSDC) and differential scanning calorimetry (DSC). The ability of TSDC to decompose a distributed relaxation, as the glass transition, into its elementary components is demonstrated. Two polarization techniques, windows polarization (WP) and non--isothermal windows polarization (NIW), are employed to assess the influence of thermal history in the results. The Tool--Narayanaswami--Moynihan (TNM) model has been used to fit the TSDC spectra. The most important contributions to the relaxation comes from modes with non--linearity ($x$) around $0.7$. Activation energies yield by this model are located around $1$~eV for polarization temperature ($T_p$) below $50$~$^\circ$C and they raise up to values higher than $8$~eV as $T_p$ increases (up to $80$~$^\circ$C). There are few differences between results obtained with WP and NIW but, nonetheless, these are discussed. The obtained kinetic parameters are tested against DSC results in several conditions. Calculated DSC curves at several cooling and heating rates can reproduce qualitatively experimental DSC results. These results also demonstrate that modelization of the non--equilibrium kinetics involved in TSDC spectroscopy is a useful experimental tool for glass transition studies in polar polymers.

\end{abstract}


\newpage

\section{Introduction}
\label{intro}

A glass is a material that behaves mechanically like a solid but has a disordered structure. In fact, glasses can be considered as overcooled liquids that have acquired solid--like rigidity. The most common way of making a glass is by cooling the material from the melt fast enough to avoid crystallization. If a liquid is cooled in such a way, molecules will rearrange progressively more slowly as viscosity increases. Eventually the structure of the material will not be able to reach the equilibrium conformation fast enough to follow the cooling rate. This falling out of equilibrium is called the glass transition \cite{kenna}. From this point of view, the glass transition would not be a true phase transition because all physical magnitudes would change in a continuous way. It has been suggested that it exists an underlying phase transition \cite{kenna} but anyway the glass transition occurs across a narrow temperature range and therefore we can define a glass transition temperature ($T_g$). Below $T_g$ only slow relaxation processes whereby the glass evolves spontaneously towards equilibrium can take place. This is known as structural relaxation or physical aging. Due to the kinetic nature of the glass transition, $T_g$ depends on the cooling rate, however this dependence turns out to be rather weak \cite{ediger}.

It is commonly accepted that the glass transition in polymers is a distributed process. In such processes, a distribution of relaxation times (DRT) is necessary to adequately describe its behavior. Calorimetric and dilatometric techniques have been widely employed to study the glass transition and several phenomenological models are used to modelize these results. In polar materials, dielectric spectroscopy techniques are also well suited to study the glass transition as the change in mobility of the main polymer chain, that occurs at $T_g$, will result in a change in the polarizability of the material. The relaxation response of a dielectric medium to an external electric field is known as dielectric relaxation. As a whole it is the result of the movement of dipoles (dipolar relaxation) and electric charges. Dynamic electrical analysis (DEA) is a well-known method that is based on the interaction of an external alternating field with the electric dipoles present in the sample. Although the aforementioned techniques are extensively employed in the study of the glass transition and related phenomena, some complexity arises in the interpretation of the results because of the distributed nature of the glass transition.

Although not as widely employed as DEA, thermally stimulated depolarization currents (TSDC) is another valuable technique to study dielectric relaxation. Its results are equivalent to those of DEA at very low frequencies ($10^{-3}$--$10^{-4}$~Hz). TSDC have shown greater sensitivity in studying the glass transition of polymers. The most interesting feature of TSDC is the ability to resolve a complex relaxation into its elementary components \cite{laca} as discussed below.

A TSDC study can be described in the following way: the sample is taken from an initial temperature ($T_i$) to a polarization temperature ($T_p$) and is polarized for a polarization time ($t_p$) by a static polarizing field ($E_p$). Then it is cooled down to the storage temperature ($T_s$) and the field is switched off at a certain point during the cooling ramp. $T_s$ is low enough so that depolarization of the sample takes place at a very low rate. By this way an electret is formed. The sample is then heated at a constant rate and the thermally activated depolarization current is recorded as a function of temperature to obtain a TSDC spectrum of the sample. Electret behavior and TSDC have been widely described in the literature \cite{sessler,bozena,chen}.

If the polarizing field, in a TSDC experiment, is on during a large portion of the cooling ramp (conventional polarization) many relaxation processes can be detected in the same thermogram. Instead, if the electric field is switched off at a point of the cooling ramp near $T_p$ or even at the beginning of the cooling ramp (windowing polarization or WP), a multiple and/or distributed spectra can be resolved into elementary spectra \cite{zielinski,sauer,laca2}. If the thermal window is narrow enough, the mode of the process that is activated will behave approximately in an elementary way and will be well described by a single relaxation time ($\tau$).

One of the relaxations that can be detected by TSDC is the $\alpha$ relaxation, which is the dielectric manifestation of the glass transition. In TSDC data, the $\alpha$ peak has its maximum at the dynamic value of $T_g$ \cite{belana}. Unlike in other relaxations \cite{johari,johari2,mudarra}, the Arrhenius law
\begin{equation}
\tau(T)=\tau_0 \exp \left(\frac{E_a}{RT}\right)
\label{arrheniuslaw}
\end{equation}
does not work properly when the system is close to the glass transition because cooperative phenomena are involved \cite{vidal}. The temperature dependence is better described by a Vogel--Tammann--Fulcher (VTF) type expression
\begin{equation}
\tau(T)= \tau_0 \exp \left[ \frac{E_w}{R(T - T_\infty)} \right] ,
\label{vogelfulcher}
\end{equation}
where $T_\infty$ is the temperature at which molecules would ``freeze'' and the relaxation time of the system would become infinite. It can be shown that this model is equivalent to the Williams--Landel--Ferry equation.

In spite of the fact that the VTF model describes the $\alpha$ relaxation better than the Arrhenius model, well known non--linear phenomena like thermal history effects and physical aging can not be explained by this model. As a consequence, the parameters obtained through the VTF model will depend to some extent on the thermal history of the sample. 

In order to modelize such memory effects, at least an additional parameter that takes into account the structural conformation of the system must be introduced. An appropriate parameter is the fictive temperature $T_f$. The fictive temperature of a non-equilibrium system is the temperature of an equilibrium system with the same structural conformation. The simplest extension to the Arrhenius model that incorporates this parameter is the Tool--Narayanaswamy--Moynihan (TNM) model \cite{moynihan}
\begin{equation}
\tau(T) = \tau_0 \exp \left(\frac{x E_a}{RT}\right) \exp \left[\frac{(1-x) E_a}{RT_f}\right].
\label{tnm}
\end{equation}
In this model the separation between $T$ and $T_f$ is introduced through a non--linearity parameter $x$ ($0 \leq x \leq 1$). The other two parameters are the pre--exponential factor $\tau_0$ and the activation energy $E_a$. The evolution of $T_f$ with time and temperature is introduced assuming an ideal-viscous return to equilibrium that depends on the relaxation time $\tau(t)$ by the equation
\begin{equation}
\frac{dT_f}{dt} = \frac{1}{\tau(t)}(T - T_f).
\label{viscosity1}
\end{equation}
Equations \ref{tnm} and \ref{viscosity1} are coupled. Their exact solution can be expressed as
\begin{equation}
T_f(t) = T(t) - \phi[z(t)] [T_f(0) - T(0)] - \int_0^t \frac{dT(t')}{dt'} \phi[z(t)-z(t')] dt'.
\label{exact}
\end{equation}
where the reduced time ($z$) is
\begin{equation}
z(t) \equiv \tau_0 \int_0^t \frac{dt'}{\tau(t')} dt'
\label{reducedtime}
\end{equation}
and the response function $\phi$ is, in the case of elementary relaxations,
\begin{equation}
\phi(t) \equiv \exp \left( - \frac{t}{\tau_0} \right).
\label{elemres}
\end{equation}
Numerical methods are required to integrate equations~\ref{exact} and \ref{reducedtime}. Nevertheless, they are often found in the literature since they can be used in some DRT implementations, like the Kohlrausch-Williams-Watts (KWW) distribution, just changing equation~\ref{elemres} for a modified response function \cite{kww}.

In this work we present a TSDC based analysis that is able to characterize structural relaxation near the glass transition. We will take advantage of possibilities of TSDC/WP to determine the DRT more clearly than with calorimetric methods. Using this analysis we will obtain an acceptable correlation with DSC data. Throughout, the similarities and differences with current approaches will be stressed. 

\section{Experimental}

\label{exp}

Experiments were carried out on commercial poly(ethylene terephthalate) (PET), Hosta-PET~\textregistered, 100$\mu$m thick sheets. Samples were almost amorphous with less than $3$\% crystallinity degree. From previous works \cite{cold} it was known that for this material the dynamic value of $T_g$ is approximately $80$~$^\circ$C at $2.5$~$^\circ$C/min and that crystallization does not take place under $100$~$^\circ$C.

Circular samples were prepared for current measurements coating $1$ cm diameter Al electrodes on both sides of the sheet by vacuum deposition. Thermally stimulated currents have been carried out in a non-commercial experimental setup, controlled by an Eurotherm--2416 temperature programmer. Temperature, during measurements, was measured to an accuracy of $0.1$~$^\circ$C by a J--thermocouple located inside the electrodes (in direct contact with the sample). A Keithley--6512 electrometer has been employed for the current intensity measurements. 

TSDC experiments were performed using the null width polarization windowing method, a particular case of the windowing polarization (WP) technique (see Section~\ref{intro}) and the non isothermal windowing polarization (NIW) technique. All the TSDC experiments begin at $T_i$ above $T_g$ to ensure that there is no influence from previous thermal history. 

In WP the sample is then cooled to the polarization temperature ($T_p$). Once the sample is at $T_p$, a polarizing potential ($V_p$) is applied after a time $t_d$. $V_p$ is applied for a polarizing time ($t_p$). Once $t_p$ is over, the polarizing potential is switched off and the sample is cooled until the storage temperature ($T_s$) is attained. The sample remains at $T_s$ for a short storage time ($t_s$) and then it is heated at a constant rate while the TSDC discharge recorded. In all WP experiments, $T_i=95$~$^\circ$C, $V_p=800$~V, $t_d=2$~min, $t_p=5$~min, $T_s=30$~$^\circ$C, $t_s=5$~min and the cooling and heating rate is $2.5$~$^\circ$C/min. 

To avoid the different thermal history in each measurement, inherent to the WP method, the NIW method was also used. In this method the sample is continuously cooled from $T_i$ to $T_s$ and the polarizing field is applied during the cooling ramp when the temperature of the sample reaches $T_p$ and switched off $\Delta T = 2$~$^\circ$C below $T_p$. As in WP experiments, in NIW experiments $T_i=95$~$^\circ$C, $V_p=800$~V, $T_s=30$~$^\circ$C, $t_s=5$~min and the cooling and heating rate is $2.5$~$^\circ$C/min.

A spectrum using conventional polarization was obtained using a polarizing potential $V=800$~V. The sample was cooled from a temperature above $T_p=95$~$^\circ$C to $T_s=30$~$^\circ$C. The electric field was on between $95$ and $40$~$^\circ$C, without isothermal polarization. This broad temperature range allows to polarize most modes of the mechanism. The sample was at $T_s$ for $5$~min and then it was heated while the TSDC was recorded. The cooling and heating rates were the same as in the other experiments ($2.5$~$^\circ$C/min).

Calorimetric measurements were made with a Mettler TC11 thermoanalyser equipped with a Mettler--20 Differential Scanning Calorimeter module. The calorimeter has been previously calibrated with metallic standards (indium, lead, zinc). DSC curves were obtained from $8$ to $20$~mg samples, sealed in aluminum pans. All DSC experiments begin at $90$~$^\circ$C, clearly above $T_g$ but low enough so crystallization does not take place. Samples are cooled to $40$~$^\circ$C and heated again to $90$~$^\circ$C. The values $1.25$~$^\circ$C/min, $2.5$~$^\circ$C/min and $5$~$^\circ$C/min have been employed for the cooling and heating rates.

\section{Results and discussion}
\label{resdis}

\subsection{Modelization procedure}
\label{gequ}

If we assume that structural relaxation at the glass transition and dielectric $\alpha$ relaxation are two manifestations of the same physical phenomenon, we can use TSDC together with WP to track each mode of the distributed relaxation independently of the other ones. This is a huge difference with other techniques such as DSC that do not have this ability. It also suggests a possible way to modelize the distribution of relaxation times that can take advantage of this feature of TSDC, although its validation should come from comparison with experimental data.

According to this approach, each mode of the structural relaxation evolves independently from the other ones. This results in as many equations \ref{tnm} and  \ref{viscosity1} as modes of the process, each one with its own fictive temperature. The relaxation time distribution is characterized by the sets $\{g_i\}$, $\{\tau_{0i}\}$, $\{E_{ai}\}$ and $\{x_i\}$. $\{g_i\}$ is a set of normalized weights that stand for the relative contribution of each mode of the relaxation. $\{\tau_{0i}\}$ and $\{E_{ai}\}$ are the pre--exponential factors and the activation energies of the different modes of the relaxation.

There are several differences between this approach and the usually employed KWW method \cite{kww}. When the KWW method is used, the relaxation is characterized only by three parameters (an activation energy $E_a$, a pre-exponential factor $\tau_0$ and a non--exponential relaxation parameter $\beta$). This method may be appropriate to modelize data obtained by techniques that can not resolve individual modes of the relaxation but more flexibility is desired to take the maximum advantage of TSDC. Another difference is that there is a single $T_f$ within the KWW method, while in the presented approach all the modes have their own $T_{fi}$. 

The intensity current that corresponds to an elementary mode of the $\alpha$ relaxation is given by
\begin{equation}
I_i(t) \propto \exp \left[ - \int_{t0}^t \frac{dt'}{\tau_i(t')} - \ln[\tau_i(t)] \right].
\label{current}
\end{equation}
In this equation, the dielectric relaxation time is considered to be the same as the structural relaxation one.

In this work, the parameters that correspond to each mode are obtained fitting a calculated depolarization current $I(T)$ obtained from equation~\ref{current} to experimental TSDC/WP and TSDC/NIW data by means of $\chi^2$ minimization. The routines used are described by other authors \cite{press}. Two different numerical approaches have been tested to obtain $\tau_i(t)$: (a) numerical resolution of equations~\ref{tnm} and \ref{viscosity1} (b) numerical integration of equation~\ref{exact}. The coincidence in the results of both methods demonstrates the validity of the numerical algorithms.

As mentioned above, validation of the assumptions that have been employed should come from comparison with experimental data. Calorimetric DSC curves can be calculated from the fictive temperature $T_{fi}(t)$ obtained for each mode during the calculation process. A general fictive temperature is defined by:
\begin{equation}
T_f(t) = \sum_{i=1}^{n} g_i T_{fi}(t)
\label{Tf(t)}
\end{equation}
and from this quantity the normalized calorific capacity is obtained by \cite{cpnorm}
\begin{equation}
C_p^n = \frac{dT_f}{dT}
\label{Cp}
\end{equation}
This result can be compared with experimental DSC scans after they are normalized.

\subsection{Analysis of TSDC data}
\label{anaexpdat}

A spectrum obtained by conventional polarization, as specified in Section~\ref{exp}, is presented in Figure~\ref{convpol}. The broad asymmetric peak corresponds to the $\alpha$ relaxation. It has a maximum at $79$~$^\circ$C. The peak contains the contribution of most modes of the $\alpha$ relaxation. It can be resolved in its elementary components using fractional polarization techniques, such as WP or NIW.

The sets $\{g_i\}$, $\{\tau_{0i}\}$, $\{E_{ai}\}$ and $\{x_{i}\}$ are obtained analyzing a set of elementary TSDC spectra that correspond to several polarization temperatures. We have employed an equispaced set of polarization temperatures from $T_p=40$~$^\circ$C up to $T_p=86$~$^\circ$C in $2$~$^\circ$C steps to obtain a representative set of parameters. The same set of polarization temperatures has been used for WP and NIW experiments.

Figure~\ref{rma1} shows a subset of these spectra, obtained by WP, corresponding to the main modes that reproduce the relaxation. The $\alpha$ dipolar peak presents its maximum value of intensity for $T_p=74$~$^\circ$C, at a temperature $T_m=80$~$^\circ$C. We will refer to this polarization temperature as the optimal polarization temperature ($T_{po}$). As $T_p$ is increased or decreased from this value, the peak maximum shifts in temperature. This indicates clearly a distributed process. Each spectra records a mode of the relaxation and the analysis of all of them allows us to study the distribution.

To obtain spectra as elementary as possible, a small polarization time has been used ($t_p=5$~min) and the electric field has been switched off just before the start of the cooling ramp. In this way we limit the contribution from modes with a relaxation time different than the one that corresponds to the main mode that is being polarized. Moreover, physical aging during the isothermal polarizing stage is minimized and as a consequence polarization is more sharply focused on a single mode. Anyway, there is the possibility that more than one mode of the relaxation process can be recorded in each measurement. To analyze if this effect appreciably affects the results, a set of TSDC spectra was obtained polarizing by the NIW method with $\Delta T=2$~$^\circ$C (see Section~\ref{exp}), using the same polarization temperatures. The thermal history of a NIW experiment is simpler, at the expense of a lower intensity signal. 

The NIW spectra are presented in Figure~\ref{rma2}. There are no appreciable differences from the spectra obtained by WP, as it would be expected if both methods are activating almost elementary spectra of the same relaxation. The overall intensity of the peaks is lower since in NIW experiments the polarizing field is on for a shorter amount of time. The $T_{po}$ is the same as in WP experiments ($T_{po}=74$~$^\circ$C) and its maximum is placed at the same temperature ($T_m=80$~$^\circ$C).

$\tau_{0i}$, $E_{ai}$ and $x_i$ for each mode can be obtained fitting numerical calculations of $I(t)$ (Equation \ref{current}) to experimental data. First, fits have been performed setting these three parameters free. Unlike $\tau_{0i}$ and $E_{ai}$, $x_i$ values did not show a broad distribution. For this reason, the mean values were calculated ($x=0.683$ for WP and $x=0.713$ for NIW) and the fittings were performed again fixing the $x$ parameter to these values and setting free only $\tau_{0i}$ and $E_{ai}$. In the following, we will omit the subindex $i$ to lighten the notation. Tables~\ref{tres} and \ref{quatre} reproduce the parameters obtained in this way for each curve. Curves corresponding to lower polarization temperatures have been neglected as their intensity is very low and, as a consequence, the evaluation of their parameters becomes uncertain. 

As a general trend, in the case of WP, we can see that $\log_{10}(\tau_{0}/1\rm{s})$ decreases as $T_p$ increases, with values ranging from $-22.8$ to $-108$. The activation energy increases as $T_p$ does so, with values from $1.70$~eV to $7.67$~eV The values obtained from the NIW spectra are comparable and show a similar behavior.

Figure~\ref{fit1} shows a comparison between the experimental curve polarized by WP at $T_p=74$~$^\circ$C and the curve calculated from the fit results. An overall good agreement can be observed. Similar plots can be obtained for the other modes, either from WP or NIW experiments. In all cases, fitting to the TNM model we are able to reproduce closely the experimental results.

The relative weight of each contribution, $g$, is roughly proportional to the polarization of the peak which, at its turn, is proportional to its area. To take into account the dependence of polarizability with temperature, the area is multiplied by $T_p$ to obtain $g$ \cite{laca}. Figure~\ref{pesos} shows the relative distribution of weights for the $\alpha$ relaxation, calculated in this way. The distributions obtained by WP and by NIW are very similar, especially when $T_p<T_{po}$.

There is a linear relationship between $\log_{10}(\tau_0)$ and $E_a$ that can be appreciated in Figure~\ref{compen}. This relationship is known as compensation law \cite{laca}. The concordance between the WP and NIW linear regression parameters is very accurate. For WP the intersect is $A=0.0930$~eV and the slope is $B=-0.0702$~eV while for NIW the intersect is $A=0.0979$~eV and the slope is $B=-0.0705$~eV. For both regression coefficients, the first five significant figures are nine.

The slight differences between WP and NIW results in Figure~\ref{pesos} gives us some clues about the merits of both polarization techniques. Although the compensation plots shown in Figure~\ref{compen} are almost identical, WP and NIW modes are on different points of the compensation line. It can be interpreted that in spite of giving almost elementary spectra, both methods do not focus on exactly the same mode for a given $T_p$. In figure~\ref{pesos} it can be seen that the reason of this discrepancy is different whether $T_p$ lies above or below $T_{po}$. 

For polarization temperatures above $T_{po}$, WP weights (Figure~\ref{pesos}) lie below the corresponding values obtained by the NIW method. This behavior can be explained if we assume that significant depolarization occurs in the WP method in the meantime while the polarizing field is switched off and the cooling ramp effectively starts. Due to the characteristic thermal inertia of the used experimental setup, roughly $30$~s pass before the cooling rate reaches $2.5$~$^\circ$C/min. This implies a considerable depolarization taking into account that the temperature of the sample is above the static $T_g$.

Therefore, weights obtained by means of WP experiments can be undervalued for these modes. NIW experiments do not present this effect because the field is applied during the cooling ramp. 

Below $T_{po}$, the weights provided by both methods agree better. They tend to lie in the same curve but at different points. Physical aging during polarization in the WP curves is the most probable cause of this trend. The reduction of free volume helps to maintain the polarization of the activated modes once the polarizing stage is over. For this reason, there is a better agreement between the weight curves obtained by means of both methods. 

It can be observed that, for these polarization temperatures, a mode with a higher $E_a$ is activated by WP, when the same $T_p$ as with NIW is employed. Again, physical aging of the sample during polarization can explain this behavior. Restriction of chain mobility during aging shifts the depolarization peak to higher temperatures \cite{canadas}, delivering higher activation energies. We can see as well that this shift in $E_a$ is more pronounced for $T_p$ around $10$~$^\circ$C below $T_{po}$ and decreases as $T_p$ approaches to $T_{po}$ as expected.

All in all, both polarization methods give comparable results when they are applied to the entire relaxation but the lack of physical aging or isothermal depolarization makes NIW results easier to understand and a little bit more reliable. For this reason, in the following lines we will center our discussion on NIW results, although WP results are similar and lead to the same conclusions in all cases.

In Figure~\ref{eatp} $T_p$ is plotted vs $E_{a}$ for NIW curves. It can be seen that the increase of $E_a$ is small for low $T_p$ and speeds up as $T_p$ reaches higher values. This trend is confirmed by more qualitative results obtained fitting the curves polarized at lower $T_p$ (presented as error bars due to uncertainty in the fittings). We can infer, then, that the activation energies of the low $E_a$ modes of the relaxation should not be less than $1$~eV.

\subsection{Comparison with DSC results}
\label{dscres}

Justification of the employed method should come from the comparison of its results with experimental data. For this purpose, calorimetric DSC curves of the glass transition were calculated from equations \ref{Tf(t)} and \ref{Cp}. The relaxation has been modelized using the parameter values presented in Table~\ref{quatre}, that correspond to eleven modes. 

Experimental plots have been normalized as follows. The tangent to the initial part of the experimental plot has been calculated and subtracted to the entire plot. Then, the experimental data has been multiplied by a constant factor in order to make equal to $1$ the high temperature end of the plot. The same normalization parameters have been employed for curves with the same heating rate.

Calculated DSC curves are plotted in Figures~\ref{pujada} and \ref{baixada}, together with the corresponding experimental results. An offset of $3.1$~$^\circ$C has been subtracted to all the temperatures of the calculated curves to fit their maxima to the experimentals ones. The more probable cause of this temperature offset may be associated to some thermal gradient in the TSDC setup, although it could also be due to limitations of the employed approach. Once the temperature offset has been corrected, we can appreciate qualitatively good agreement in both cases. In Figure~\ref{pujada} we can see the effect of changing the cooling rate. There is a higher enthalpy recovery for lower cooling rates, that is reproduced by the model. In Figure~\ref{baixada} the effect of changing the heating rate is shown. In this case the enthalpy recovery is more pronounced for higher heating rates. This behaviour is also reproduced by the model. Dependence of the temperature where the transition occurs with the change in the heating or cooling rate is reproduced by the model in both cases.

In Figure~\ref{comparacio} the experimental and the calculated DSC curves are compared side by side. Some differences between them can be observed. The experimental endothermic peak, once normalized, is less pronounced than the predicted one. This effect could be related to thermal inertia in the DSC sample. Also, the inclusion of low $E_a$ modes in the calculation would lead to a better concordance in the part of the curve that corresponds to the glassy state.

\section{Conclusions}

It has been known for a long time that the $\alpha$ relaxation is the dielectric manifestation of the glass transition \cite{belana}. Following this idea, in this work we suggest a method to study the glass transition using TSDC and we show how to correlate these results with DSC data.

TSDC together with WP or NIW can be used to analyze the different modes of the DRT on its own. It has been assumed that each mode of the DRT evolves according to its own $T_f$. This hypothesis allows us to obtain acceptable results.

DSC curves calculated from parameters obtained by the study of the TSDC spectra show a fair agreement with experimental DSC curves and reproduce the qualitative trends observed when the kinetics of the experiment is changed. Quantitative differences may be due to several reasons that have been discussed in the text.

We have employed two different polarization methods. A comparative analysis between results obtained by WP and NIW polarization has lead us to employ preferentially results obtained from NIW experiments. There are no major differences between results obtained by either method but the simpler thermal history of NIW makes its results somewhat more reliable.

TSDC is a method that should be taken into account to study the glass transition. Its main advantage is its capability to study elementary modes of the relaxation. The high resolution of the TSDC spectra is another reason to consider this technique. To sum up, TSDC is a technique worth considering to study the glass transition in polar polymers.

\bibliography{petrma}

\begin{thebibliography}{10}
\newcommand{\enquote}[1]{``#1''}

\bibitem{kenna}
G.~B. McKenna, {\em Comprehensive Polymer Science\/}, volume 2 (Polymer
  Properties), chapter~10, pp. 311--358, Pergamon, Oxford (1990).

\bibitem{ediger}
M.~D. Ediger, C.~A. Angell and S.~R. Nagel, {\em J. Phys. Chem.\/} {\bf 100},
  13200--13212 (1996).

\bibitem{laca}
A.~B. G.~Teyss\`edre, S.~Mezghani and C.~Lacabanne, {\em Dielectric
  Spectroscopy of Polymeric Materials\/}, chapter~8, pp. 227--258, American
  Chemical Society, Washington, {DC} (1997).

\bibitem{sessler}
G.~M. Sessler, {\em Electrets\/}, volume~2, chapter~10, pp. 41--80, Laplacian,
  Morgan Hill, {CA}, 3rd edition (1999).

\bibitem{bozena}
B.~Hilczer and J.~Malecki, {\em Electrets\/}, chapter~6, pp. 285--313, Studies
  in electrical and Electronical Engineering 14, Elsevier--PWN--Polish
  Scientific, Warszawa (1986).

\bibitem{chen}
R.~Chen and Y.~Kirsh, {\em Analysis of Thermally Stimulated Processes\/},
  chapter~3, pp. 60--81, Pergamon, Oxford, 1st edition (1981).

\bibitem{zielinski}
M.~Zielinski and M.~Kryszewski, {\em Phys. Stat. Sol. A\/} {\bf 42}, 305--314
  (1977).

\bibitem{sauer}
B.~B. Sauer and P.~Avakian, {\em Polym.\/} {\bf 33}, 5128--5142 (1992).

\bibitem{laca2}
G.~Teyss\`edre and C.~Lacabanne, {\em J. Phys. D: Appl. Phys.\/} {\bf 28},
  1478--1487 (1995).

\bibitem{belana}
J.~Belana, P.~Colomer, S.~Montserrat and M.~Pujal, {\em Macromol. Sci. Phys.
  B\/} {\bf 23(4-6)}, 467--481 (1984--1985).

\bibitem{johari}
G.~P. Johari and M.~Goldstein, {\em J. Chem. Phys.\/} {\bf 53}, 2372--2388
  (1970).

\bibitem{johari2}
G.~P. Johari, {\em J. Chem. Phys.\/} {\bf 58}, 1766--1770 (1973).

\bibitem{mudarra}
M.~Mudarra and J.~Belana, {\em Polym.\/} {\bf 38}, 5815--5821 (1997).

\bibitem{vidal}
E.~V. Russell and N.~E. Israeloff, {\em Nature\/} {\bf 408}, 695--698 (2000).

\bibitem{moynihan}
C.~T. Moynihan, A.~J. Easteal, J.~Wilder and J.~Tucker, {\em J. Phys. Chem.\/}
  {\bf 78}, 2673--2677 (1974).

\bibitem{kww}
G.~Williams and D.~C. Watts, {\em Trans. Faraday Soc.\/} {\bf 66}, 80--85
  (1970).

\bibitem{cold}
J.~C. Ca{\~n}adas, J.~A. Diego, J.~Sellar\`es, M.~Mudarra and J.~Belana, {\em
  Polym.\/} {\bf 41}, 8393--8400 (2000).

\bibitem{press}
W.~H. Press, B.~P. Flannery, S.~A. Teukolsky and W.~T. Vetterling, {\em
  Numerical Recipes\/}, chapter~10, pp. 387--448, Cambridge University Press,
  Cambridge, 2nd edition (1992).

\bibitem{cpnorm}
I.~M. Hodge and A.~R. Berens, {\em Macromolecules\/} {\bf 14}, 1598--1599
  (1981).

\bibitem{canadas}
J.~C. Ca{\~n}adas, J.~A. Diego, M.~Mudarra and J.~Belana, {\em Polym.\/} {\bf
  39}, 2795--2801 (1998).

\end{thebibliography}

\newpage

\pagestyle{empty}

\begin{table}[h]

\caption{Kinetic parameters obtained in the fitting process of TSDC/WP spectra to the TNM model. $x=0.683$ for every mode. \label{tres}}
\begin{center}

\begin{tabular}{ccccc}\hline
& $T_p$~($^\circ$C) & $\log_{10}(\tau_{0}/1\rm{s})$ & $E_a$~(eV) & $g/g_{max}$ \\ \hline 
a & $60$ & $-22.8$ & $1.70$ & $0.298$ \\
b & $62$ & $-25.5$ & $1.89$ & $0.379$ \\
c & $64$ & $-30.1$ & $2.21$ & $0.470$ \\
d & $66$ & $-35.8$ & $2.61$ & $0.579$ \\
e & $68$ & $-43.0$ & $3.12$ & $0.713$ \\
f & $70$ & $-50.6$ & $3.65$ & $0.856$ \\
g & $72$ & $-60.5$ & $4.34$ & $0.977$ \\
h & $74$ & $-70.8$ & $5.07$ & $1.00$ \\
i & $76$ & $-80.8$ & $5.77$ & $0.794$ \\
j & $78$ & $-93.8$ & $6.70$ & $0.330$ \\
k & $80$ & $-108$ & $7.67$ & $0.0442$ \\
\end{tabular}

\end{center}

\end{table}

\newpage

\begin{table}[h]

\caption{Kinetic parameters obtained in the fitting process of TSDC/NIW spectra to the TNM model. $x=0.713$ for every mode. \label{quatre}}
\begin{center}

\begin{tabular}{ccccc}\hline
& $T_p$~($^\circ$C) & $\log_{10}(\tau_{0}/1\rm{s})$ & $E_a$~(eV) & $g/g_{max}$ \\ \hline 
a & $60$ & $-20.2$ & $1.53$ & $0.235$ \\
b & $62$ & $-23.5$ & $1.76$ & $0.301$ \\
c & $64$ & $-27.8$ & $2.06$ & $0.390$ \\
d & $66$ & $-33.2$ & $2.43$ & $0.504$ \\
e & $68$ & $-40.1$ & $2.92$ & $0.652$ \\
f & $70$ & $-48.8$ & $3.52$ & $0.820$ \\
g & $72$ & $-58.7$ & $4.22$ & $0.973$ \\
h & $74$ & $-71.6$ & $5.12$ & $1.00$ \\
i & $76$ & $-85.1$ & $6.08$ & $0.762$ \\
j & $78$ & $-98.1$ & $7.00$ & $0.330$ \\
k & $80$ & $-111$ & $7.93$ & $0.0472$ \\
\end{tabular}

\end{center}

\end{table}

\clearpage

\newtheorem{peudefigura}{Figure}

\begin{peudefigura}
{\rm TSDC spectra of the $\alpha$ relaxation of PET obtained by conventional polarization.}
\label{convpol}
\end{peudefigura}

\begin{peudefigura}
{\rm TSDC spectra of the $\alpha$ relaxation of PET obtained by WP at different $T_p$: from $60$~$^\circ$C (curve a) to $80$~$^\circ$C (curve k) in $2$~$^\circ$C increment steps.}
\label{rma1}
\end{peudefigura}

\begin{peudefigura}
{\rm TSDC spectra of the $\alpha$ relaxation of PET obtained by NIW at different $T_p$: from $60$~$^\circ$C (curve a) to $80$~$^\circ$C (curve k) in $2$~$^\circ$C increment steps.}
\label{rma2}
\end{peudefigura}

\begin{peudefigura}
{\rm Calculated (continuous) and experimental (circles) TSDC spectrum of PET for $T_p=74$~$^\circ$C (Fitted values: $\tau_0=1.46 \times 10^{-71}$~s, $E_a=5.07$~eV, $x=0.683$).}
\label{fit1}
\end{peudefigura}

\begin{peudefigura}
{\rm Obtained distribution of weights $g_i$ for each analyzed mode of the $\alpha$ relaxation. Polarization technique: WP (circle, continuous line), NIW (triangle, dashed line).}
\label{pesos}
\end{peudefigura}

\begin{peudefigura}
{\rm Compensation plot: WP (circle, dashed line), NIW (triangle, dashed line).}
\label{compen}
\end{peudefigura}

\begin{peudefigura}
{\rm Relation between $E_a$ and $T_p$ for NIW curves. The dashed line is a guide for the eye.}
\label{eatp}
\end{peudefigura}

\begin{peudefigura}
{\rm $C_p^{n}$ for $2.5$~$^\circ$C/min heating rate. Cooling rate is $1.25$~$^\circ$C/min (continuous), $2.5$~$^\circ$C/min (dashed) and $5$~$^\circ$C/min (point--dashed). (a)~Experimental DSC curves. (b)~Calculated curves.}
\label{pujada}
\end{peudefigura}

\begin{peudefigura}
{\rm $C_p^{n}$ for $2.5$~$^\circ$C/min cooling rate. Heating rate is $1.25$~$^\circ$C/min (continuous), $2.5$~$^\circ$C/min (dashed) and $5$~$^\circ$C/min (point--dashed). (a)~Experimental DSC curves. (b)~Calculated curves.}
\label{baixada}
\end{peudefigura}

\begin{peudefigura}
{\rm Comparison of $C_p^{n}$ obtained by DSC (continuous) and by calculation (dashed). Cooling and heating rate $2.5$~$^\circ$C/min.}
\label{comparacio}
\end{peudefigura}

\clearpage

\newcommand{\dibuix}[2]{%

\newpage

\pagestyle{empty}

\hbox{}\vspace{2cm}

\begin{center}
\includegraphics[width=10cm]{#1}
\end{center}

\vspace{3cm}

\noindent
{\bf Figure #2}

\noindent
J. A. Diego {\em et al.}, ``Dielectric study \ldots''.

}

\dibuix{complet}{\ref{convpol}}

\dibuix{TSC00252_3}{\ref{rma1}}

\dibuix{TSC00251_2}{\ref{rma2}}

\dibuix{ajuste}{\ref{fit1}}

\dibuix{Im_vs_Ea_3}{\ref{pesos}}

\dibuix{compensation}{\ref{compen}}

\dibuix{Tp_vs_Ea_3}{\ref{eatp}}

\dibuix{pujada250-4}{\ref{pujada}}

\dibuix{baixada250-4}{\ref{baixada}}

\dibuix{comparativa-250-250-4}{\ref{comparacio}}


\end{document}